\documentclass[12pt]{article}
\usepackage{graphicx}
\usepackage{amssymb}
\usepackage{amsmath}
\numberwithin{equation}{section}

\newtheorem{Pa}{Paper}[section]
\newtheorem{Tm}[Pa]{{\bf Theorem}}
\newtheorem{La}[Pa]{{\bf Lemma}}
\newtheorem{Cy}[Pa]{{\bf Corollary}}
\newtheorem{Rk}[Pa]{{\bf Remark}}

\newtheorem{Ee}[Pa]{{\bf Example}}
\newtheorem{Dn}[Pa]{{\bf Definition}}

\title{Relativistic Lippmann - Schwinger equation}

\author{Lev Sakhnovich}

\date{}

\begin{document}

\maketitle

\emph{99 Cove ave. Milford, CT 06461, USA,}\\

 E-mail: lsakhnovich@gmail.com\\

 

\begin{abstract}The classical Lippmann-Schwinger equation plays an important role in the
scattering theory (non-relativistic case, Schr\"odinger   equation). In the present paper we consider the relativistic analogue of the Lippmann-Schwinger equation. We represent the corresponding  equation
in the integral form. Using this integral equation we investigate the stationary scattering problems
 (relativistic case, Dirac equation).
 We consider the dynamical scattering problems (relativistic case, Dirac equation) as well. \end{abstract}

\section{Introduction}The classical integral
Lippmann-Schwinger equation plays an important role in the  scattering theory (non-relativistic case,  Schr\"odinger   equation).
 The relativistic analogue of the Lippmann-Schwinger equation was formulated in the terms of the limit
values of the corresponding resolvent. In the present paper we found the limit values of the resolvent
in the explicit form. Using this result, we represent relativistic Lippmann-Schwinger equation (RLS equation) as an integral equation (Sections 2 and 3). 
In Section 4, we consider the dynamical scattering problems (relativistic case, Dirac equation).
In Section 5, we show that the integral RLS equation is effective by investigating
the stationary scattering problems (relativistic case, Dirac equation).

It is interesting to compare the results of dynamic and stationary
scattering theory. The corresponding results for the radial case were obtained in \cite{Sakh1} and \cite{Sakh3}.
\section{RLS equation in the integral form}
1. Let us write the Dirac equation (see \cite{BetSalp})
\begin{equation}i\frac{\partial}{\partial{t}}u(r,t)=\mathcal{L} u(r,t),\label{2.1}\end{equation}
where $u(r,t)$ is $4\times1$ vector function and $r=(r_1,r_2,r_3)$. The operators $\mathcal{L}$ and $\mathcal{L}_{0}$ are defined by the relations
\begin{equation}\mathcal{L}u=[-e\nu(r)I_{4}+m\beta+{\alpha}({p}+e{A}(r))]u,\quad
\mathcal{L}_{0}u=(m\beta+{\alpha}{p})u. \label{2.2}\end{equation}
Here $p=-i$ grad, $\nu$ is a scalar potential, $A$ is a vector potential, $(-e)$ is the electron charge.
 Now let us define $\alpha=[\alpha_1,\alpha_2,\alpha_3]$. The matrices $\alpha_k$  are the $4{\times}4$ matrices of the forms
\begin{equation}\alpha_k=\left(
                           \begin{array}{cc}
                             0 & \sigma_k \\
                             \sigma_k & 0 \\
                           \end{array}
                         \right), \quad k=1,2,3,\label{2.3}\end{equation}
where
\begin{equation}\sigma_1=\left(
                           \begin{array}{cc}
                             0 & 1 \\
                             1 & 0 \\
                           \end{array}
                         \right),\quad
\sigma_2=\left(
                           \begin{array}{cc}
                             0 & -i \\
                             i & 0 \\
                           \end{array}
                         \right),\quad
\sigma_3=\left(
                           \begin{array}{cc}
                             1 & 0 \\
                             0 & -1 \\
                           \end{array}
                         \right).\label{2.4}\end{equation}
The matrices $\beta$ and $I_{2}$ are defined by the relations
\begin{equation} \beta=\left(
                         \begin{array}{cc}
                            I_2 & 0 \\
                           0 & -I_2 \\
                         \end{array}
                       \right),\quad
I_2= \left(
                         \begin{array}{cc}
                          1 & 0 \\
                           0 & 1 \\
                         \end{array}
                       \right).\label{2.5}\end{equation}
2. We consider separately the unperturbed Dirac equation \eqref{2.1}, \eqref{2.2}, when $\nu(r)=0$ and $A(r)=0$. The Fourier transform is defined by
\begin{equation} \Phi(q)=Fu(r)=(2\pi)^{-3/2}\int_{R^3}e^{iqr}u(r)dr. \label{2.6}\end{equation}The inverse Fourier transform has the form
\begin{equation} u(r)=F^{-1}\Phi(q)=(2\pi)^{-3/2}\int_{Q^3}e^{-iqr}\Phi(q)dq.\label{2.7}\end{equation}
In the momentum space the unperturbed  Dirac equation takes the form (see \cite{AB}, Ch.IV):
 \begin{equation}i\frac{\partial}{\partial{t}}\Phi(q,t)=H_{0}(q)\Phi(q,t),\quad q=(q_1,q_2,q_3),\label{2.8}\end{equation}
 where $H_{0}(q)$ and $\Phi(q,t)$ are matrix functions of order $4{\times}4$ and  $4{\times}1$
respectively. Here the matrix $H_{0}(q)$ is defined by the relation
\begin{equation} H_{0}(q)=\left[ \begin{array}{cccc}
                          m & 0 & q_3 & q_1-iq_2 \\
                          0 & m & q_1+iq_2  & -q_3 \\
                          q_3 & q_1-iq_2  & -m & 0 \\
                          q_1+iq_2  & -q_3 & 0 & -m
                        \end{array}\right] .
\label{2.9}\end{equation}
The eigenvalues $\lambda_{k}$ and the corresponding eigenvectors
$g_k$ of  $H_{0}(q)$ are important in our theory.  We find them below:
\begin{equation}
\lambda_{1,2}=-\sqrt{m^2+|q|^{2}},\quad \lambda_{3,4}=\sqrt{m^2+|q|^{2}}
\quad ( |q|^2:=q_{1}^2+q_{2}^2+q_{3}^2);
\label{2.10}\end{equation}
\begin{equation}
g_1=\begin{bmatrix}(-q_1+iq_2)/(m+\lambda_3) \\ q_3/(m+\lambda_3) \\ 0 \\ 1\end{bmatrix}, \quad
g_2=\begin{bmatrix}-q_3/(m+\lambda_3) \\ (-q_1- iq_2)/(m+\lambda_3) \\1 \\ 0 \end{bmatrix},
\label{2.11}
\end{equation}
\begin{equation}g_3=\begin{bmatrix} (-q_1+iq_2)/(m-\lambda_3)\\ q_3/(m-\lambda_3)\\ 0 \\ 1\end{bmatrix},
\quad
g_4= \begin{bmatrix} -q_3/(m-\lambda_3) \\ (-q_1-iq_2)/(m-\lambda_3) \\ 1 \\ 0\end{bmatrix}.
\label{2.12}\end{equation}
It follows from \eqref{2.9} and \eqref{2.10} that
\begin{equation}H_{0}^{2}(q)=(m^2+|q|^2)I_4.\label{2.13}\end{equation}
Hence we obtain
\begin{equation}H_{0}^{-1}(q)=(m^2+|q|^2)^{-1}H_{0}(q).\label{2.14}\end{equation}
Let $|\lambda|<|\lambda_1(|q|)|$. Using \eqref{2.13} we have
\begin{equation}(H_{0}(q)-\lambda)^{-1}=H_{0}^{-1}(q)+H_{0}^{-1}(q)\frac{\lambda^2}
{\lambda_{1}^{2}(|q|)-\lambda^{2}}+\frac{\lambda}{\lambda_{1}^{2}(|q|)-\lambda^{2}}.\label{2.15}\end{equation}
In view of analyticity of both parts of equality \eqref{2.15} this equality is valid at all
$\lambda{\notin}E,$ where $E=(-\infty,-m]{\bigcup}[m,+\infty)$.\\
\begin{Cy}\label{Corollarry 2.1}The operator $\mathcal{L}_{0}$ has no eigenvalues in the interval $(-m.m)$.\end{Cy}
3. \emph{Now we will construct a relativistic analogue of the Lippmann-Schwinger equation
(RLS integral equation).}\\
To do it we consider the expression
\begin{equation}B_{\pm}(r,\lambda)=F^{-1}[H_{0}(q)-(\lambda{\pm}i0)]^{-1},\label{2.16}\end{equation}
where  $\lambda=\overline{\lambda},\, |\lambda|>m$. Let us write
the following relation (see \cite{BrPr}, formula 721).
\begin{equation}J_1(r)=F^{-1}[(m^2+|q|^2)^{-1}]= m^{1/2}K_{1/2}(m|r|)/|r|^{1/2},
\label{2.17}\end{equation}
where $K_{p}(z)$ is the modified Bessel function. It is known that (see \cite{BatErd})
\begin{equation}K_{1/2}(z)=K_{-1/2}(z)=\sqrt{\frac{\pi}{2z}}e^{-z}.\label{2.18}\end{equation}
According to \eqref{2.17} and \eqref{2.18} the equality
\begin{equation}J_1(r)=\sqrt{\pi/2}(e^{-m|r|}/|r|)\label{2.19}\end{equation}
is valid.
Using \eqref{2.19} we obtain
\begin{equation}J_2(r)=F^{-1} (\frac{q_{k}}{m^2+|q|^2})=i\frac{\partial}{\partial{r_k}}J_1(r)=
-\sqrt{\pi/2}e^{-m|r|}\frac{r_k}{|r|^{2}}(m+1/|r|).
\label{2.20}\end{equation}
Let us calculate the expression
\begin{equation}J_{\pm}(r,\lambda)=F^{-1}[\frac{1}{\lambda_{1}^{2}(q)-(\lambda{\pm}i0)^2}],\quad
\lambda{\in}E.\label{2.21}\end{equation}
In view of \eqref{2.10} and \eqref{2.21} we have
\begin{equation}J_{\pm}(r,\lambda)=-\overline{F\frac{1}{m_{1}^2-|q|^2{\mp}i(sgn{\lambda})0}},\quad
\lambda{\in}E.
\label{2.22}\end{equation}  where $m
_{1}^{2}=\lambda^{2}-m^{2},\, m_{1}>0.$  Taking into account formulas  7.19
and 7.20 from the  book \cite{BrPr} (table of the Fourier transformation) and relation
\eqref{2.18} we obtain the equalities
\begin{equation}J_{\pm}(r,\lambda)=\sqrt{\frac{\pi}{2}}e^{{\pm}im_{1}|r|}/|r|,\quad \lambda>m,\label{2.23}\end{equation}
\begin{equation}J_{\pm}(r,\lambda)=\sqrt{\frac{\pi}{2}}e^{{\mp}im_{1}|r|}/|r|,\quad \lambda<-m.\label{2.24}\end{equation}
Formulas \eqref{2.14}, \eqref{2.19} and \eqref{2.20} imply that
\begin{equation} Q(r)=F^{-1}H_{0}^{-1}(q)=\sqrt{\pi/2}e^{-m|r|}[m\beta-(m+1/|r|)r\alpha/|r|]/|r|.\label{2.25}\end{equation}
Here $r\alpha=r_1\alpha_1+r_2\alpha_2+r_3\alpha_3$, matrices $\alpha_k$ and $\beta$ are defined by the relations  \eqref{2.3}-\eqref{2.5}.
Due to \eqref{2.15}, \eqref{2.16}, \eqref{2.20} and \eqref{2.25} we have
\begin{equation}  B_{\pm}(r,\lambda)= Q(r)+(2\pi)^{3/2}\lambda^{2}Q(r)\ast{J_{\pm}(r,\lambda)}+
{\lambda}J{\pm}(r,\lambda),\label{2.26}\end{equation}
where  $ F(r)\ast{G(r)}=\int_{R^{3}}F(r-v)G(v)dv$ is the convolution of $F(r)$ and $G(r)$.
Now we can write the equation
\begin{equation}\phi_{\pm}(r,k,n)=e^{ikr}g_{n}(k)-(2\pi)^{-3/2}\int_{R^{3}}B_{\pm}(r-s,\lambda)V(s)\phi_{\pm}(s,k,n)ds,
\label{2.27}\end{equation} where
\begin{equation}V(r)=-e{\nu}(r)I_4+e{\alpha}A(r).\label{2.28}\end{equation}
Here the vectors $g_{n}(k)$ are defined by the relations \eqref{2.11} and \eqref{2.12}. \\
\emph{ Equation \eqref{2.27} (RLS equation) is relativistic analogue of the Lippmann-Schwinger equation.}\\  We note that   the
Lippmann-Schwinger equation play an important role  in the non-relativistic scattering theory (see \cite{RS}). Our aim
is to show that the constructed RLS integral equation  can be effective by solving relativistic scattering problems.
\section{Propeties of the RLS integral equation}.
1. Further we assume that the matrix $V(r)$ is self-adjoint,
\begin{equation}V(r)=V^{\star}(r).\label{3.1}\end{equation}
 Hence $V(r)$ can be represented in the form
\begin{equation}V(r)=U(r)D(r)U^{\star}(r),\label{3.2}\end{equation}
where $U(r)$ is an unitary matrix, $D(r)$ is a diagonal matrix
\begin{equation}D(r)=diag(d_{1}(r),d_{2}(r),d_{3}(r),d_{4}(r)).\label{3.3}\end{equation}
Let us introduce the diagonal matrices
\begin{equation}D_{1}(r)=diag(|d_{1}(r)|^{1/2},|d_{2}(r)|^{1/2},|d_{3}(r)|^{1/2},|d_{4}(r)|^{1/2})
\label{3.4}\end{equation}
and
\begin{equation}W(r)=diag(signd_{1}(r),signd_{2}(r),signd_{3}(r),signd_{4}(r)).\label{3.5}\end{equation}
Formulas \eqref{3.2}-\eqref{3.5} imply that
\begin{equation}V(r)=V_{1}(r)W_{1}(r)V_{1}(r),\label{3.6}\end{equation}
where
\begin{equation}V_{1}(r)=U(r)D_{1}(r)U^{\star}(r),\quad W_{1}(r)=U(r)W(r)U^{\star}(r)\label{3.7}\end{equation}
It is easy to see that
\begin{equation}\|V_{1}(r)\|^{2}=\|V(r)\|,\quad \|W_{1}(r)\|=1.\label{3.8}\end{equation}
2. \emph{Modificated RLS integral equation.}\\
If $\phi_{\pm}(r,k,n)$ is a solution of RLS equation, then the vector-function $\psi_{\pm}(r,k,n)=V_{1}(r)\phi_{\pm}(r,k,n)$
is a solution of following  modified RLS integral equation:
\begin{equation}\psi_{\pm}(r,k,n)=e^{ikr}V_{1}(r)g_{n}(k)-(2\pi)^{-3/2}B_{\pm}(\lambda)\psi_{\pm}(r,k,n),
\label{3.9}\end{equation}
where
\begin{equation}B_{\pm}(\lambda)f=\int_{R^{3}}V_{1}(r)B_{\pm}(r-s,\lambda)V_{1}(s)W_{1}(s)f(s)ds.
\label{3.10}\end{equation} We note that the operators $B_{\pm}(\lambda)$ act   in the Hilbert space
$L_{4}^{2}(R^{3})$ of $4{\times}1$ vector functions.
\begin{Tm}\label{Theorem 3.1}Let condition \eqref{3.1} be fulfilled,
the function  $\|V(r)\|$ be bounded and belong to the space $L^{1}(R^{3}).$
Then the operators $B_{\pm}(\lambda)$ are compact.
\end{Tm}
\emph{Proof.} We represent the operators $B_{\pm}(\lambda)$ in the form
\begin{equation}B_{\pm}(\lambda)=\sum_{m=1}^{3}B_{\pm}(m,\lambda),\label{3.11}\end{equation}
where
\begin{equation}B_{\pm}(m,\lambda)f=\int_{R^{3}}V_{1}(r)B_{\pm}(r-s,m,\lambda)V_{1}(s)W_{1}(s)f(s)ds.
\label{3.12}\end{equation}
Here the $4{\times}4$ matrix functions $B_{\pm}(r,m,\lambda)$ are defined by the relations
\begin{equation}B_{\pm}(r,1,\lambda)=
{\lambda}J_{\pm}(r,\lambda),\label{3.13}\end{equation}
\begin{equation}B_{\pm}(r,2)=Q(r)\label{3.14}\end{equation}
\begin{equation}B_{\pm}(r,3,\lambda)=(2\pi)^{3/2}\lambda^{2}Q(r){\ast}J_{\pm}(r,\lambda).
\label{3.15}\end{equation}
Formulas (2.23), \eqref{2.24} and \eqref{3.13} imply
that
\begin{equation}\|B_{\pm}(r,1,\lambda)\|{\leq}C(\lambda)/|r|.\label{3.16}\end{equation}
According to condition of the theorem the function $\|V(r)\|$ belongs to the Rolnik class (see \cite{RS}), i.e.
\begin{equation}\int_{R^3}\int_{R^3}\frac{\|V(r)\|\|V(s)\|}{|r-s|^{2}}dsdr<\infty.\label{3.17}\end{equation}
It follows from \eqref{3.12}, \eqref{3.16} and \eqref{3.17} that the operator $B_{\pm}(1,\lambda)$ belongs to the Hilbert-Schmidt class. Hence operator $B_{\pm}(1,\lambda)$ is compact.\\
Let us consider the operator $B_{\pm}(2)$.
In view of \eqref{2.25}   we have
\begin{equation}C_1=\int_{R^3}\|Q(r)\|dr<\infty.\label{3.18}\end{equation}
Hence the operator $B_{\pm}(2)$ is bounded  (see \cite{Sakh2}, section 1.4)and
\begin{equation}\|B_{\pm}(2)\|{\leq}MC_1,\quad M=sup\|V(r)\|.\label{3.19}\end{equation}
We represent the kernel $B_{\pm}(r,2)$ in the form $B_{\pm}(r,2)=B_{\pm}(r,2,1)+B_{\pm}(r,2,2)$ where
\begin{equation}B_{\pm}(r,2,1)=Q(r),\, 0<r<\epsilon,\quad B_{\pm}(r,2,1)=0, \, r>\epsilon,\label{3.20}\end{equation}
\begin{equation}B_{\pm}(r,2,2)=0,\, 0<r<\epsilon,\quad B_{\pm}(r,2,2)=Q(r), r>\epsilon,\label{3.21}\end{equation}
We introduce the operators
\begin{equation}B_{\pm}(2,m)f=\int_{R^{3}}V_{1}(r)B_{\pm}(r-s,2,m)V_{1}(s)W_{1}(s)f(s)ds,\quad m=1,2.
\label{3.22}\end{equation}
It is easy to see, that the operator $B_{\pm}(2,2)$ belongs to the Hilbert= Schmidt class and
\begin{equation}\|B_{\pm}(2)-B_{\pm}(2,2)\|=\|B_{\pm}(2,1)\|{\leq}M\int_{0}^{\epsilon}\|Q(r)\|dr.\label{3.23}
\end{equation}The norm $\|B_{\pm}(2,1)\|$  tends to zero when $\epsilon {\to}0.$
Hence, it follows from \eqref{3.23} that the operator $B_{\pm}(2)$ is compact.\\
To consider the operator $B_{\pm}(3,\lambda)$ we use the inequality
\begin{equation}\|B_{\pm}(r,\lambda,3)\|{\leq}C(\lambda)\frac{e^{-m|r|}}{|r|}{\ast}(\frac{1}{|r|}+\frac{1}{|r|^2}).
\label{3.24}\end{equation}
It follows from \eqref{3.24} and Adams theorem (see Appendix, Examples 6.2 and 6.3), that
\begin{equation}\int_{R^3}\int_{R^3}\|V(r)\|\|B_{\pm}(r-s,3,\lambda)\|^{2}\|V(s)\|dsdr{<}\infty. \label{3.25}
\end{equation}
According to \eqref{3.25} the operator $B_{\pm}(3,\lambda)$  belongs to the Hilbert-Schmidt class.
Thus, all the operators $B_{\pm}(m,\lambda),\,(m=1,2,3)$ are compact. The theorem is proved.\\
\section{Wave  and scattering operators,dynamical case}
We introduce the operator function
\begin{equation}\Theta(t)=exp(it\mathcal{L})exp(-it\mathcal{L}_{0}).
\label{4.1}\end{equation}
The wave operators $W_{\pm}(\mathcal{L},\mathcal{L}_0)$ are defined by the relation (see \cite{RS}).
\begin{equation}W_{\pm}(\mathcal{L},\mathcal{L}_0)=\lim_{t{\to}\pm{\infty}}\Theta(t)P_0.\label{4.2}
\end{equation} Here  $P_0$ is orthogonal projector
on the absolutely continuous subspace $G_0$ with respect to the operator $\mathcal{L}_0$. The
limit in \eqref{4.2} supposed to be  in the sense of strong convergence.
\begin{Tm}\label{Theorem 4.1} If $V(r)=V^{*}(r)$,
the function  $\|V(r)\|$ is bounded, belongs to the space $L^{1}(R^{3})$ and
\begin{equation}\int_{-\infty}^{+\infty}[\int_{|r|>|t|\epsilon}\|V(r)\|^{2}dr]^{1/2}dt<\infty,\quad \epsilon>0, \label{4.3}\end{equation}
then the wave operators $W_{\pm}(\mathcal{L},\mathcal{L}_0)$ exist.
\end{Tm}
\emph{Proof.} We use the equality
\begin{equation}\Theta(t)-I=\int_{0}^{t}\frac{d}{dt}\Theta(t)dt.\label{4.4}\end{equation}
Thus, to prove the formulated theorem, it is sufficient to show that the following inequality
holds
\begin{equation}\int_{-\infty}^{+\infty}\|\frac{d\Theta(t)}{dt}\Psi\|{dt}<\infty
 \label{4.5}\end{equation}
on a set $S$  vector functions $\Psi$ dense in  $L_{4}^{2}(R^3)$.
We consider the vector functions
\begin{equation}\tilde{\Psi}(p)=\frac{1}{(2\pi)^{3/2}}\int_{R^3}
\Psi(r)e^{-ipr}dr.\label{4.6}\end{equation}
We take such a set $S$ of vector functions  $\Psi(r)$  that the corresponding vector functions  $\tilde{\Psi}(p)$ belong to the class $C^{\infty}$ and
\begin{equation}supp{\tilde{\Psi}(p)}{\in}\{\|p\|:0<c_1(\Psi)<\|p\|<c_2(\Psi)<\infty\}.
\label{4.7}\end{equation}
Let us  consider the case, when $t>0$. We have
\begin{equation}\|\frac{d\Theta(t)\Psi}{dt}\|=\|J(r,t)\|.\quad \label{4.8}\end{equation}
where $\Psi(r)$ belongs to the set $S$ and
\begin{equation}J(r,t)=(2\pi)^{-3/2}\int_{R^3}V(r)exp(-itH_{0}(p))e^{ipr}\tilde{\Psi}(p)dp.
\label{4.9}\end{equation}
Using \eqref{4.9}
we obtain
\begin{equation}\|\frac{d\Theta(t)}{dt}{\Psi}(r)(\|{\leq}(\|J_{1}^{2}(r,t)\|)^{1/2}+(\|J_{2}^{2}(r,t)\|)^{1/2}
\label{4.10}\end{equation}
where
\begin{equation}J_{k}(r,t)=\int_{R^3}{V(r)}
exp[itF_{k}(p,r,t)]\tilde{\Psi_{k}}(p){dp},\quad k=1,2.\label{4.11}\end{equation}
Here
\begin{equation}F_{k}(p,r,t)=pr/t-\mu_{k},\quad \mu_{1}=-\mu_2=\sqrt{|p|^{2}+m^2}.\label{4.12}\end{equation}
We note that $\mu_1(p)$ and $\mu_2(p)$ are the eigenvalues of the matrix $H_{0}(p)$, the vectors
$\tilde{\Psi}_{k}(p)$ are the corresponding eigenvectors.
The stationary-phase points $p_{k}(r,t)$ are the solutions of the equations
\begin{equation}\frac{\partial}{{\partial}p_{s}}F_{k}(p,r,t)=0,\quad s=1,2,3.
\label{4.13}\end{equation}
Thus, we have
\begin{equation}r-tp/{\mu_{k}}=0.
\label{4.14}\end{equation}
When $\epsilon>0$ is small the stationary-phase points do not belong to the region $\|r\|{\leq}t\epsilon$. Hence integrating by parts the right side of \eqref{4.11} we have
(see \cite{Fed}):
\begin{equation}\int_{\|r\|t{\leq}\epsilon}(\|J_{k}(r,t)\|)^{2}dr{\leq}ct^{-4},\quad k=1,2.\label{4.15}\end{equation}
We use here the relation $\|V(r)\|^{2}{\in}L(R^3)$.\\
Now let us consider the case when $\|r\|{\geq}t\epsilon.$
It follows from \eqref{4.11}  that
\begin{equation}\|J_{k}(r,t)\|{\leq}c\|V(r)\|.\label{4.16}\end{equation}
Taking into account the condition \eqref{4.3} of the theorem we obtain
\begin{equation}
\int_{-\infty}^{+\infty}[\int_{\|r\|{\geq}t\epsilon}\|J_{k}(r,t)\|^{2}dr]^{1/2}dt<\infty
\label{4.17}\end{equation}
The relations \eqref{4.10}, \eqref{4.15} and \eqref{4.17} imply inequality \eqref{4.5}.
The theorem is proved.
\begin{Rk}\label{Remark 4.2}Let condition \eqref{3.1} be fulfilled. If
the function  $\|V(r)\|$ is bounded and
\begin{equation}\|V(r)\|{\leq}\frac{M}{|r|^{\alpha}},\quad |r|{\geq}\delta>0,\quad \alpha>3,
\label{4.18}\end{equation}
then the wave operators $W_{\pm}(\mathcal{L},\mathcal{L}_0)$ exist.
\end{Rk}
\begin{Dn}\label{Definition 4.3}The scattering operator $S_{\pm}(\mathcal{L},\mathcal{L}_0)$ is defined by the relation
\begin{equation}S(\mathcal{L},\mathcal{L}_0)=W_{-}^{*}(\mathcal{L},\mathcal{L}_0)W_{+}(\mathcal{L},\mathcal{L}_0).
\label{4.19}\end{equation}\end{Dn}
\section{Stationary scattering problem}
1.In section 4 we considered the dynamical scattering problem $(t{\to}\infty)$
for Dirac equation \eqref{2.1}, \eqref{2.2}. In the present section we shall investigate the stationary
scattering problem  for the same equation. It means that we shall investigate
 asymptotic behavior of $\psi_{+}(r,k,n)$ when $|r|{\to}\infty$.\\
 (The case $\psi_{-}(r,k,n)$ can be investigated similarly).
\begin{Dn}\label{Definition 5.1}We say that $\lambda{\in}E$ is an exceptional value if the equation $[I+(2\pi)^{-3/2}B_{+}(\lambda)]\psi=0$ has nontrivial solution in the space $L^2(R^3)$.\end{Dn}
 We denote by $\mathcal{E}_{+}$ the set of exceptional points and we denote by $E_{+}$ the set
 of such points $\lambda$ that $\lambda{\in}E,\quad \lambda{\notin}\mathcal{E}_{+}.$\\
 Using Theorem 3.1 and the Fredholm alternative we obtain
\begin{La}\label{Lemma 5.2}Let conditions of Theorem 3.1 be fulfilled. If $\lambda{\in}E_{+}$, then equation \eqref{3.9} has one and only one
solution $\psi_{+}(r,k,n)$ in $L^2(R^3)$. \end{La}
\begin{Cy}\label{Corollary 5.3}Let conditions of Theorem 3.1 be fulfilled. If $\lambda{\in}E_{+}$
, then equation \eqref{2.27} has one and only one
solution $\phi_{+}(r,k,n)$ which satisfies the condition $V_{1}(r)\phi_{+}(r,k,n){\in}L^2(R^3)$. \end{Cy}
It can be
 proved  (see \cite{RS},XI, III) that
\begin{La}\label{Lemma 5.4}Let conditions of Theorem 3.1 be fulfilled. The set $\mathcal{E}_{+}$ is   closed and has  Lebesgue  measure equal to zero. \end{La}
2.\emph{ Let us consider the case $V(r)=0$ separately.}\\
Taking into account \eqref{2.26} we see that $B(r,\mu)$ is defined when $\Im{\mu}>0$:
\begin{equation}
  B_{+}(r,\mu)= Q(r)+(2\pi)^{3/2}\mu^{2}Q(r)\ast{J_{+}(r,\mu)}+
{\lambda}J_{+}(r,\mu),\label{5.1}\end{equation},
where
\begin{equation}J_{+}(r,\mu)=\sqrt{\frac{\pi}{2}}e^{im_{1}|r|}/|r|,\quad m_{1}(\mu)=\sqrt{\mu^2-m^2},
\quad \Im{\mu}>0.\label{5.2}\end{equation}
We assume that $\Im{m_{1}}(\mu)>0.$
 We introduce the operator
\begin{equation}{B}_{0}(\mu)f=\int_{R^3}B_{+}(r-s,\mu)f(s)ds,\label{5.3} \end{equation}
where $f(r){\in}L^2(R^3)$. It is easy to see that the operator ${B}_{0}(\mu)$
is bounded in the space $L^2(R^3)$.
It follows from \eqref{5.3} that
\begin{equation}F[{B}_{0}(\mu)f]=(2\pi)^{3/2}F[{B}_{0}(\mu)]F(f)
\label{5.4}\end{equation}
We represent the equality \eqref{5.4} in the form
\begin{equation}{B}_{0}(\mu)f=(2\pi)^{3/2}F^{-1}\{F[{B}_{0}(\mu)]F\}F^{-1}[F(f)]
\label{5.5}\end{equation}
Relation \eqref{2.2} and \eqref{2.16}  imply:
\begin{equation}F^{-1}\{F[{B}_{0}(\mu)]F\}=(\mathcal{L}_{0}-\mu)^{-1}.\label{5.6}\end{equation}
It follows from \eqref{5.4} and \eqref{5.6} that
\begin{equation}(\mathcal{L}_{0}-\mu)^{-1}f=(2\pi)^{-3/2}\int_{R^3}B_{+}(r-s,\mu)f(s)ds,\quad \Im{\mu}>0.\label{5.7}\end{equation}
3.\emph{Now we prove the main result of this section.}\\
\begin{Tm}\label{Theorem 5.5}Let the $V(r)=V^*(r)$ and the function $\|V(r)\|$ be bounded and belong to the space $L^{1}(R^3)$. If
\begin{equation}\|V_1(r)\|=O(|r|^{-3/2}),\quad |r|{\to}\infty,\label{5.8}\end{equation}
then the solution $\phi_{+}(r,k,n)$ of RLS equation \eqref{2.27} has the form
\begin{equation}\phi_{+}(r,k,n)=e^{ikr}g_{n}(k)+\frac{e^{im_{1}(k)|r|}}{|r|}f(\omega,k,n)
+o(1/|r|),\quad |r|{\to}\infty , \label{5.9}\end{equation},
where $\lambda{\in}E_{+}, \, |k|^2=\lambda^2-m^2,\,\omega=r/|r|$ and
\begin{equation}f(\omega,k,n)=-\frac{1}{4\pi}{\lambda}\int_{R^3}e^{-im_{1}(k)s\omega}
V(s)\phi_{+}(s,k,n)ds.  \label{5.10}\end{equation}
 \end{Tm}
\emph{Proof.} The equation \eqref{3.9} has one and only one solution $\psi_{+}(r,k,n)$ in the space $L^{2}(R^3)$. We shall estimate the integral
\begin{equation}J=\int_{R^3}B_{+}(r-s,\lambda)V_{1}(s)W_{1}(s)\psi_{+}(s,k,n)ds.\label{5.11}\end{equation}
We divide the space $R^{3}$ with respect to $s$ into three parts:\\
$\mathcal{M}_{1}=\{|s|{\leq}|r|/2\},\quad \mathcal{M}_{2}=\{|r|/2{\leq}|s|{\leq}3|r|/2\},\quad \mathcal{M}_{3}=\{|s|{\geq}3|r|/2\}.$\\
Let us  introduce the integrals
\begin{equation}J_{p,m}=\int_{\mathcal{M}_{p}}B_{+}(r-s,\lambda,m)V_{1}(s)W_{1}(s)\psi_{+}(s,k,n)ds,
\label{5.12}\end{equation} where $1{\leq}p{\leq}3,\quad 1{\leq}m{\leq}3.$ Taking into account relations
\eqref{2.23}-\eqref{2.25} and \eqref{3.13},\eqref{3.14} we have
\begin{equation}\|B_{+}(r,\lambda,m)\|=O(1/|r|),\quad |r|{\to}\infty,\quad m=1,2.\label{5.13}\end{equation}
Then the following relations
\begin{equation}|J_{1,1}|+|J_{1,2}|=O(1/|r|),\quad |J_{3,1}|+|J_{3,2}|=o(1/|r|),\quad |r|{\to}\infty\label{5.13'}\end{equation} are valid. We have used here the inequality
\begin{equation}\int_{R^3}\|\psi_{+}(s,k,n)\|^{2}ds<\infty.\label{5.14}\end{equation}
Condition \eqref{5.8} of the Theorem 5.5 and relations \eqref{5.11}, \eqref{5.14} imply
\begin{equation}(|J_{2,1}|+|J_{2,2}|)=o(|r|^{-3/2})(\int_{\mathcal{M}_{2}}\frac{ds}{|r-s|^2})^{1/2}=o(|r|^{-1}),
\quad |r|{\to}\infty.\label{5.15}\end{equation}
Using \eqref{3.15} and  spaces $\mathcal{M}_{p}$ we obtain
\begin{equation}\|B_{+}(r,\lambda,3)\|=O(1/|r|),\quad |r|{\to}\infty.\label{5.16}\end{equation}
Now in the same way as in cases m=1 and m=2 we receive:
\begin{equation}|J_{1,3}|=O(1/|r|),\quad |J_{2,3}|+|J_{3,3}|=o(1/|r|),\quad |r|{\to}\infty\label{5.17}\end{equation}
It is easy to see that
\begin{equation} |J_{1,2}|+|J_{1,3}|=o(1/|r|),\quad |r|{\to}\infty\label{5.18}\end{equation}
So, we have proved the relation
\begin{equation} J(r,\lambda,n)|=J_{1,1}|+o(1/|r|),\quad |r|{\to}\infty\label{5.19}\end{equation}
We note that
$|r-s|^2=|r|^2-2rs+|s|^2$. Hence we have
\begin{equation}|r-s|{\sim}|r|-s\omega,\quad \omega=r/|r|.\label{5.20}\end{equation}
The assertion of the Theorem 5.5 follows directly from \eqref{5.19} and \eqref{5.20}.
\begin{Dn}\label{Definition 5.6} The $4{\times}1$ vector function $f(\omega,k,n)$ we name the relativistic scattering
amplitude.\end{Dn}
We note that relativistic scattering amplitude is defined by formulas \eqref{5.9} and \eqref{5.10}
which are similar to the
corresponding
formulas  for non-relativistic  scattering amplitude (see \cite{RS}).\\
4. Now we shall investigate the connection between solutions of the equation
\begin{equation}\mathcal{L}\phi=\lambda\phi\label{5.22}\end{equation}
and the solutions of the  RLS  equation \eqref{2.27}.
\begin{Tm}\label{Theorem 5.7} Let the vector function $\phi(r)$ satisfies the equation \eqref{5.22} and the following conditions are fulfilled\\
$\psi=V_{1}\phi{\in}L^{2}_{4}(R^3)$  and  $\kappa_{Q}\phi{\in}L^{2}_{4}(R^3)$,\\
where $\lambda{\in}E$ and $\kappa_{Q}$ is the characteristic function of a  bounded domain $Q$.\\
If the matrix function $V(r)$ satisfies the conditions of the Theorem 3.1, then the vector  function $\psi(r)$ satisfies the equation
\begin{equation}\psi(r)=-(2\pi)^{-3/2}B_{+}(\lambda)\psi(r),\quad \lambda{\in}E.\label{5.23}\end{equation}\end{Tm}
\emph{Proof.} We use the equality
\begin{equation}z(\lambda+i\epsilon)\phi=V_{1}(\lambda+i\epsilon-\mathcal{L}_{0})^{-1}(\lambda+i\epsilon-\mathcal{L})\phi=
[I+(2\pi)^{-3/2}B_{+}(\lambda+i\epsilon)]\psi,
\label{5.24}\end{equation} where  $\epsilon>0.$
Taking into account   \eqref{5.7} and \eqref{5.22}, we obtain:
\begin{equation}\kappa_{Q}z(\lambda+i\epsilon)\phi=
i\epsilon(\kappa_{Q}V_{1})(\lambda+i\epsilon-\mathcal{L}_{0})^{-1}\phi=i\epsilon(\kappa_{Q}V_{1})B_{0}(\lambda+i\epsilon)\phi
\label{5.25}\end{equation}
According to \eqref{5.1}-\eqref{5.3} the operator $(\kappa_{Q}V_{1})B_{0}(\lambda+i\epsilon)$
belongs to the Hilbert-Schmidt class with norm
\begin{equation}\|(\kappa_{Q}V_{1})B_{0}(\lambda+i\epsilon)\|=O(\epsilon^{-1/2}),\quad \epsilon{\to}0.
\label{5.26}\end{equation}
We have
\begin{equation}\kappa_{Q}z(\lambda+i\epsilon)\phi{\to}0,\quad \epsilon{\to}0.\label{5.27}\end{equation}
The assertion of the Theorem 5.7 follows directly from \eqref{5.24}and \eqref{5.27}
\begin{Cy}\label{Corollary 5.8} Let the conditions of the Theorem 5.7 be fulfilled.
If $\lambda{\in}E$ is eigenvalue of the corresponding operator $\mathcal{L}$, then $\lambda{\in}\mathcal{E}$.\end{Cy}
\begin{Tm}\label{Theorem 5.9}Let the conditions of Theorem 5.5 be fulfilled and let  the function $\phi_{+}(r,k,n)$ be a solution of RLS equation \eqref{2.27} such that $V_{1}(r)\phi_{+}(r,k,n){\in}L^{2}((R^3).$ If $\lambda{\in}E_{+}$ then the function $\phi_{+}(r,k,n)$
is the solution of the equation \eqref{5.22}
in the distributive sense.
\end{Tm}
\emph{Proof.} We consider the expression
\begin{equation}(\mathcal{L}_{0}-\lambda)(\mathcal{L}_{0}-\lambda-i\epsilon)^{-1}V\phi_{+}=
V\phi_{+}+i\epsilon(\mathcal{L}_{0}-\lambda-i\epsilon)^{-1}V\phi_{+}.\label{5.28}\end{equation}
In the same way as \eqref{5.27} we prove that
\begin{equation}\kappa_{Q}\epsilon\|(\mathcal{L}_{0}-\lambda-i\epsilon)^{-1}V_{1}\|{\to}0,\quad
\epsilon{\to}0.\label{5.29}\end{equation}
 Using relation \eqref{2.27} we have
\begin{equation}(\mathcal{L}_{0}-\lambda-i\epsilon)^{-1}V\phi_{+}=-\phi_{+}+o(1),\quad \epsilon{\to}0.
\label{5.30}\end{equation}
Now we introduce the class of functions f(r) such that $f(r){\in}C_{\infty}$ and $f(r)=0$ when $r{\notin}Q$. Taking into account \eqref{5.29} we obtain
\begin{equation}\lim_{\epsilon{\to}0}((\mathcal{L}_{0}-\lambda)(\mathcal{L}_{0}-\lambda-i\epsilon)^{-1}V\phi_{+},f)=
(V\phi_{+},f)=(\phi_{+},Vf).\label{5.31}\end{equation}
According to  \eqref{5.30} the equality
\begin{equation}\lim_{\epsilon{\to}0}((\mathcal{L}_{0}-\lambda)(\mathcal{L}_{0}-\lambda-i\epsilon)^{-1}V\phi_{+},f)=
-(\phi_{+},(\mathcal{L}_{0}-\lambda)f)\label{5.32}\end{equation} holds. It follows from \eqref{5.31} and \eqref{5.32} that
\begin{equation}(\phi_{+},\mathcal{L}f)=0.\label{5.33}\end{equation}
Relation  \eqref{5.33} implies the assertion of the Theorem 5.9.
\section{Appendix}
1. In this section we will consider the integral
\begin{equation}U(r)=\int_{R^3}\frac{f(s)}{|r-s|^{\lambda}}ds.\label{6.1}\end{equation}
Let us formulate a partial case of Adams theorem (see \cite{SL}).\\
\begin{Tm}\label{Theorem 6.1} We assume that $1<p<q,\, \lambda>3/p^{\prime}$, where $1/p+1/p^{\prime}=1.$
If the condition
\begin{equation}
\sup_{\rho>0}[\rho^{3/p^{\prime}-\lambda+3/q}]<\infty \label{6.2}\end{equation}
is fulfilled,
then
\begin{equation}\|U\|_{L_q}=[\int_{R^3}|U(r)|^{q}dr]^{1/q}{\leq}c\|f\|_{L_p},\label{6.3}\end{equation}
where the constant $c$  is independent of $f$.\end{Tm}
\begin{Ee}\label{Example 6.2} We consider the case when $\lambda=1, \,  p^{\prime}>3.$ \end{Ee}
According to Theorem 6.1 we have
\begin{equation}1<p=\frac{p^{\prime}}{p^{\prime}-1}<3/2, \quad q=\frac{3p^{\prime}}{p^{\prime}-3}>3>p,
\quad U(r){\in}L_{q}(R^3).
\label{6.4}\end{equation}
\begin{Ee}\label{Example 6.3} We consider the case when $\lambda=2, \, 3<p^{\prime}<6.$ \end{Ee}
According to Theorem 6.1 we have
\begin{equation}1<p=\frac{p^{\prime}}{p^{\prime}-1}<3/2, \quad q=\frac{3p^{\prime}}{2p^{\prime}-3}>2>p,
\quad U(r){\in}L_{q}(R^3).\label{6.5}\end{equation}
2. Using Examples 6.2 and 6.3 we will prove the inequality \eqref{3.25}. To do it we introduce the functions
\begin{equation}U_{k}(r)=\frac{e^{-m|r|}}{|r|}{\ast}[\frac{1}{|r|^{k}},\quad k=1,2.
\label{6.6}\end{equation}
We have
\begin{equation}\frac{e^{-m|r|}}{|r|}{\in}L_{p}(R^3),\quad 1<p<3/2.\label{6.7}\end{equation}
Hence,
Example 6.2 implies that
\begin{equation}U_{1}(r){\in}L_q(R^3),\quad U_{1}^{2}(r){\in}L_{q/2}(R^3),\quad q>3.
\label{6.8}\end{equation}
It follows from the  conditions of Theorem 3.1 that $\|V(r)\|$ belongs to $L_{p}(R^3)$ for all $1{\leq}p<\infty.$ Then we obtain
\begin{equation}\int_{R^3}\int_{R^3}\|V(r)\||U_{1}(r-s)|^{2}\|V(s)\|dsdr{<}\infty. \label{6.9}
\end{equation}
Now we consider the function $U_{2}(r)$.
We have
\begin{equation}\frac{e^{-m|r|}}{|r|^{2}}{\in}L_{p}(R^3),\quad 1<p<3/2.\label{6.10}\end{equation}
Hence,
Example 6.3 implies that
\begin{equation}U_{2}(r){\in}L_q(R^3),\quad U_{2}^{2}(r){\in}L_{q/2}(R^3),\quad q>2.
\label{6.11}\end{equation}
Then we obtain
\begin{equation}\int_{R^3}\int_{R^3}\|V(r)\||U_{2}(r-s)|^{2}\|V(s)\|dsdr{<}\infty. \label{6.12}
\end{equation} The inequality \eqref{3.25} follows directly from \eqref{6.9} and \eqref{6.12}.

\end{document}